\documentclass[aps,prc,twocolumn,superscriptaddress,nofootinbib,
preprintnumbers,showkeys,floatfix,tightenlines]{revtex4-2}
\usepackage[utf8]{inputenc}
\usepackage[english]{babel}
\usepackage{amssymb,amsthm,amsmath,amstext,amsbsy,amsopn}
\usepackage{bbm}
\usepackage{nicefrac}
\usepackage{slashed}
\usepackage{physics}
\usepackage{siunitx}
\usepackage{graphicx}
\usepackage{hyperref}
\usepackage[capitalise]{cleveref}
\usepackage{leftidx}
\usepackage{environ}
\usepackage{mathtools}
\usepackage{xspace}
\usepackage{array}
\usepackage{xcolor}
\usepackage{pgfplotstable}
\usepackage{booktabs}
\usepackage{float}
\usepackage{placeins}
\usepackage{isotope}
\usepackage{tikz}
\usepackage{orcidlink}
\usepackage{comment}

\AtBeginDocument{\RenewCommandCopy\qty\SI} 

\usetikzlibrary{shapes,patterns,decorations.markings}

\newcommand{\pwl}{l}

\newcommand{\ii}{\mathrm{i}}

\NewEnviron{subalign}[1][]{%
\begin{subequations}\begin{align}
  \BODY
\end{align}\label{#1}\end{subequations}
}

\NewEnviron{spliteq}{%
\begin{equation}\begin{split}
  \BODY
\end{split}\end{equation}
}

\renewcommand{\vec}[1]{\mathbf{#1}}
\newcommand{\SW}{$S$} 
\newcommand{\PW}{$P$} 
\newcommand{\DW}{$D$} 
\definecolor{ccolor}{named}{red}

\begin{document}

\title{Short-distance production of three particles with large scattering length}

\author{T.~G.~Backert\orcidlink{0000-0002-5701-0280}}
\email{timothy\_george.backert@tu-darmstadt.de}
\affiliation{Technische Universität Darmstadt, Department of Physics, Institut für
  Kernphysik, 64289 Darmstadt, Germany}

\author{S.~Dietz\orcidlink{0000-0001-5129-5476
}}\xspace 
\affiliation{Technische Universität Darmstadt, Department of Physics, Institut für
  Kernphysik, 64289 Darmstadt, Germany}

\author{H.-W.~Hammer\orcidlink{0000-0002-2318-0644}}
\email{hans-werner.hammer@physik.tu-darmstadt.de}
\affiliation{Technische Universität Darmstadt, Department of Physics, Institut für
  Kernphysik, 64289 Darmstadt, Germany}
\affiliation{ExtreMe Matter Institute EMMI and Helmholtz Forschungsakademie
  Hessen f\"ur FAIR (HFHF),
GSI Helmholtzzentrum für Schwerionenforschung GmbH,
64291 Darmstadt, Germany}

\author{S.~König\orcidlink{0000-0002-4954-0548}}
\email{skoenig@ncsu.edu}
\affiliation{Department of Physics and Astronomy, North Carolina State University,
Raleigh, NC 27695, USA}
\affiliation{Technische Universität Darmstadt, Department of Physics, Institut für
  Kernphysik, 64289 Darmstadt, Germany}

\author{D.~T.~Son\orcidlink{0000-0003-1662-4777}}
\email{dtson@uchicago.edu}
\affiliation{Leinweber Institute for Theoretical Physics,
University of Chicago, Chicago, IL 60637, USA}
\date{\today}

\begin{abstract}
The short-distance production of multiparticle states in high-energy nuclear reactions
provides a unique way to study the low-energy properties of few-body systems.
In particular, the production amplitude 
of multineutron systems is strongly constrained by an approximate conformal symmetry of
the underlying theory. We calculate the full amplitude for the short-distance production of three particles
with large scattering length in leading order pionless EFT, focusing on the cases of
three neutrons and three spinless bosons. We investigate the signature of low-energy resonances and other correlations in the
relative energy distributions.
For the case of neutrons, we compare to the predictions from approximate conformal
symmetry close to the unitary limit and calculate the range corrections up to next-to-next-to-leading order.
\end{abstract}
\maketitle

\section{Introduction}\label{sec:Intro}

Short-distance particle production provides a powerful tool for probing the interactions
of the produced particles. 
At ALICE, for example, the femtoscopy method is used to extract low-energy scattering
parameters from the correlation functions measured in heavy ion
collisions~\cite{Janik:2018ghw}.
In this paper, we investigate the short-distance production of multiparticle states,
focusing on the cases of
three neutrons and three spinless bosons.
The short-distance production of neutrons can be realized experimentally in high-energy nuclear reactions, assuming factorization of the production cross 
section~\cite{Bodwin:1994jh,Hammer:2021zxb}. Such reactions provide a method to study
multineutron systems~\cite{Duer:2022ehf}.

The study of multineutron systems has a long 
and venerable history~\cite{Kezerashvili:2016ucn,Marques:2021mqf}.
In addition to their fundamental importance in nuclear physics,
multineutron systems also provide a window on nuclear many-body
forces with extreme isospin. 
This topic has received renewed interest  
with several experiments looking for tetraneutron clusters.
Marqués and collaborators observed six events that exhibit the characteristics
of a multineutron cluster liberated in the proton-induced breakup of
\isotope[14]{Be} by detecting the recoiling proton~\cite{PhysRevC.65.044006}.
Subsequently, a few events
pointing toward a $4n$ resonance where measured in the reaction
$^4$He($^8$He,$^8$Be)$4n$~\cite{Kisamori:2016jie}.
Another experiment using the reaction $^7$Li($^7$Li,$^{10}$C) even suggested
that the tetraneutron could be bound \cite{Faestermann:2022meh}.
The most convincing evidence for a resonance-like structure to date with much
higher statistics, however, was recently presented by Duer \textit{et al.} in
an experiment using the reaction $^8$He$(p,p\alpha)4n$~\cite{Duer:2022ehf}.
Whether the resonance-like structure seen in the experiment of
Ref.~\cite{Duer:2022ehf} is a genuine resonance or due to some
other effect, such as the final state interaction among dineutrons and/or the
presence of preexisting four-neutron correlations from the $^8$He nucleus 
projectile~\cite{Lazauskas:2022mvq} remains an open question.
A nonstandard hypothesis that is able to reconcile all experimental findings
was proposed in Ref.~\cite{faestermann:2025recent}.
The central observable in hard knock-out reactions, such as
$\isotope[8]{He}(p,p\alpha)4n$, is the relative energy (or center-of-mass energy)
distribution of
the neutrons, i.e., the energy distribution in the reference frame moving with
the center-of-mass of the neutrons. 
In such a high-energy reaction, the neutrons are created in a very
small spatial region in the order of 2.5~fm -- roughly the size of $\isotope[8]{He}$~\cite{PhysRevLett.99.252501,alkhazov1997nuclear}.
Typically, most of the 
kinetic energy transferred to the neutrons in the hard knock-out 
reaction is carried by the motion of their center of mass.
This
ensures that the dominant final state interaction after the knock-out
is only among the neutrons, which have a low relative energy.

In this paper, we investigate the qualitative properties of such a multineutron state 
created at short distances in pionless effective field theory (EFT)~\cite{vanKolck:1997ut,Kaplan:1998tg,Kaplan:1998we,vanKolck:1998bw}.
We focus on the production of three neutrons in a point
and leave the extension to four neutrons and the effects of an 
extended source distribution for future work.
Similar calculations for the case of three neutral charm mesons can be found in 
Refs.~\cite{Braaten:2021iot,Braaten:2023acw}.
The point-production is calculated applying effective potentials derived from
pionless EFT at leading order (LO) (see Ref.~\cite{Dietz:2021haj} for details)
and next-to-next-to-leading order (N$^2$LO). 

A particular focus is placed on the explicit
verification of the constraints from nonrelativistic conformal 
symmetry on the energy distribution of the neutrons~\cite{Hammer:2021zxb}.
For relative energies \(E\) in the range
\(1/(ma^2) \approx \qty{0.1}{\mega\electronvolt} \ll E \ll 1/(mr^2) \approx \qty{5}
{\mega\electronvolt}\), where \(a\) and \(r\) are the scattering length and effective
range of neutrons with mass $m=939.565$~MeV, respectively, the system is effectively scale invariant.
This universal regime is known as the unitary (or unitarity) limit and plays an
important role in many strongly interacting systems ranging from ultracold atoms to
hadrons and nuclei~\cite{Braaten:2004rn,Zwerger:2012,Hammer:2017tjm,Hammer:2019poc}. 
In this regime the energy distribution is governed by power laws with exponents
determined by the scaling dimensions of certain operators in the nonrelativistic
conformal field theory for the unitary limit.
In addition, there is a second
scale invariant region for very small energies, $E\ll 1/(ma^2)$, where the exponents
are given by the scaling dimensions of the non-interacting theory.
In this region, the energy spectrum simply follows the pure phase-space behavior.
Both regions are connected to renormalization group fixed points of the coupling
constant that determines the strength of the two-body contact interaction between 
neutrons~\cite{Kaplan:1998we}.
Pionless EFT describes
the full energy spectrum up to $E \lesssim 1/(mr^2)$ including the transition between
the two scale-invariant regions.
A related study focusing on the observability of the conformal exponents was presented
in Ref.~\cite{higgins:2025observability}.

For comparison, we also consider the point-production of three spinless bosons, where 
genuine three-body resonances are known to 
exist~\cite{Dietz:2021haj,Bringas:2004zz,Deltuva:2020sdd,Hyodo:2013zxa,Jonsell:2006xx}.
In the negative scattering region, the Efimov bound states turn into resonances as they cross the three-body threshold as the scattering length is decreased. Thus they can be viewed as precursors to Efimov states. 
In this case, the continuous scale symmetry is broken down to a subgroup of discrete
scaling transformations for three or more particles.
As a consequence, physical observables show
discrete scale invariance and Efimov physics~\cite{Efimov:1970zz, Braaten:2004rn}.

This paper is organized as follows.
In Sec.~\ref{sec:pprod}, we introduce the formalism required to calculate the
point-production amplitude in pionless EFT.
Our results are presented and discussed in 
Sec.~\ref{sec:results}. 
In Sec.~\ref{sec:4comparison}, we compare our results against existing theoretical and
experimental results.
The conclusion and an outlook are given in Sec.~\ref{sec:conc}.
Since the corresponding integral equations are very similar, the three-neutron and
three-boson cases are treated together unless noted otherwise.

\section{Point-production amplitude}
\label{sec:pprod}

Our aim is to calculate the relative energy distribution of the final state
particles (neutrons or bosons), \(R(E)\), 
up to a normalization factor that is regarded as arbitrary in this work.
The distribution can be obtained by solving 
Faddeev equations for the point-production amplitude.
For this, we follow the strategy of Refs.~\cite{Platter:2004he,Platter:2004zs}
and use pionless EFT 
to construct an effective interaction potential as a sum of $n$-body potentials,
\begin{equation}
  {V}_\mathrm{eff} = \sum_{n=2}^\infty {V}_n \,.
\end{equation}

In general, interaction terms up to $n=N$ contribute in a $N$-body
problem, but at low energies higher-body terms are typically suppressed (this is described
naturally by dimensional analysis in the EFT, along with certain well-known exceptions; see, e.g.,
Ref.~\cite{Hammer:2019poc} for a review).
The potentials ${V}_n$
are constrained by Galilean symmetry and can be expressed in a momentum expansion.
For \( S \)-wave two-body interactions, we have
\begin{equation}\label{eq:2BInt}
  \langle \vec{k'}|{V}_2|\vec{k}\rangle =
  C_0 + C_{2}(\vec{k'}^2 + \vec{k}^2)/2 + \ldots \,,
\end{equation}
where $\vec{k}$ and $\vec{k'}$ are the relative momenta in the
initial and final state.
Regulator functions and spin-projection operators have been suppressed.
Similar expansions can be written down
for three- and higher-body interactions. 
In a multineutron system,
only the momentum-independent two-body contact interaction $C_0$ in the $^1S_0$
channel contributes at leading order in pionless EFT~\cite{Bedaque:1997qi,Bedaque:1998mb}.
In the three-boson system, in contrast, both a momentum-independent two-body and
three-body contact interaction have to be included to properly
renormalize the system~\cite{Bedaque:1998kg, Bedaque:1998km}.
Assuming typical momenta of order $1/a$, the uncertainty of a leading-order
pionless EFT calculation can be estimated as $|r/a|$, where $r$ is the effective
range and $a$ the scattering length.
For the three-neutron system with
$a\approx {-}18.9$ fm and
$r\approx 2.83$~fm~\cite{Gardestig:2009ya,babenko2016study,malone2022measurement},
this leads to a theoretical uncertainty in the order of 15\% at leading 
order -- see Appendix~\ref{sec:effective range corrections}.
The exact form of the effective potential, of course, depends on the specific
regularization scheme used.
However, low-energy observables are independent of the regularization
scheme (up to higher-order corrections), and one can choose any convenient
scheme for practical calculations.
In this work, we consider two different schemes, a soft Gaussian regulator and a
sharp cutoff on momentum loop integrals.

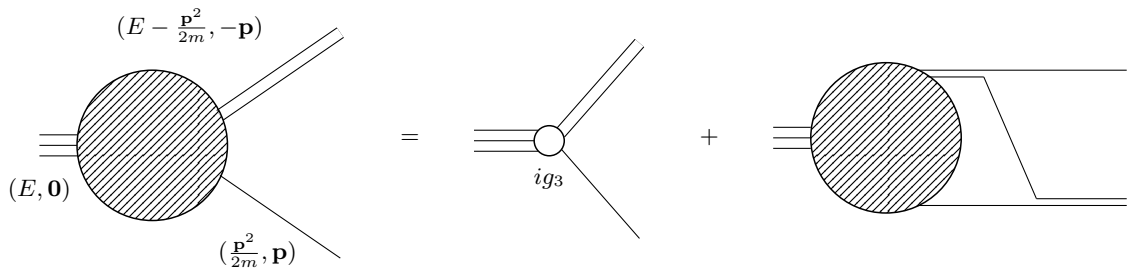
\begin{figure*}[htb]
	\begin{center}
			\begin{tikzpicture}[baseline={([yshift=-.5ex]current bounding box.center)}]
				\node (origin) at (0,0) {};
				\node () at (-1.5, -0.6) {\((E,\vec{0})\)};				
				\node () at (0.5, 1.6) {\((E -\frac{\vec{p}^2}{2m},-\vec{p})\)};			
				\node () at (1.4, -1.4) {\((\frac{\vec{p}^2}{2m},\vec{p})\)};			
				
				\draw  [] (-1.5,0.14) -- (-1.,0.14);
				\draw  [] (-1.5,0.0) -- (-1.,0.0);
				\draw  [] (-1.5,-0.14) -- (-1.,-0.14);
			
							
				\draw  [double, double distance=1.4mm] (0.9,0.4) -- (2.5,1.5);
				\draw  [] (0.9,-0.4) -- (2.5,-1.5);
                \node [circle, draw=black, fill=white,inner sep=0mm,minimum size=20mm, line width=0.2mm, pattern=north east lines] () at(origin) {};		
				\node [circle, draw=white, fill=white,inner sep=0mm,minimum size=5mm, line width=0.2mm] () at(origin) {};	
				\node () at (0,0) {$\Gamma$};
		\end{tikzpicture}
		\hspace{0.5cm}
		=
		\hspace{0.5cm}
		\begin{tikzpicture}[baseline={([yshift=-.5ex]current bounding box.center)}]
				\node (origin) at (0,0) {};
				
				\draw  [] (-1.,0.14) -- (-0.15,0.14);
				\draw  [] (-1.,0.0) -- (-0.15,0.0);
				\draw  [] (-1.,-0.14) -- (-0.15,-0.14);
							
				\draw  [double, double distance=1.4mm] (0.145,0.1) -- (1.2,1.3);
				\draw  [] (0.15,-0.1) -- (1.2,-1.3);
				\node [circle, draw=black, fill=white,inner sep=0mm,minimum size=5mm, line width=0.2mm] () at(origin) {};		
                \node () at (0.0,0.0) {\(g_3\)};
		\end{tikzpicture} 
		\hspace{0.5cm}
		+
		\hspace{0.5cm}
		\begin{tikzpicture}[baseline={([yshift=-.5ex]current bounding box.center)}]
				\node (origin) at (0,0) {};
				
				\draw  [] (-1.5,0.14) -- (-1.,0.14);
				\draw  [] (-1.5,0.0) -- (-1.,0.0);
				\draw  [] (-1.5,-0.14) -- (-1.,-0.14);
				

				\draw  [] (0.43,0.9) -- (3.2,0.9);
				\draw  [] (0.56,0.81) -- (1.3,0.81);
				\draw  [] (1.3,0.81) -- (2.0,-0.81);
				\draw  [] (2.0,-0.81) -- (3.2,-0.81);
				\draw  [] (0.43,-0.9) -- (3.2,-0.9);	
                \node [circle, draw=black, fill=white,inner sep=0mm,minimum size=20mm, line width=0.2mm, pattern=north east lines] () at(origin) {};	
				\node [circle, draw=white, fill=white,inner sep=0mm,minimum size=5mm, line width=0.2mm] () at(origin) {};	
				\node () at (0,0) {$\Gamma$};
		\end{tikzpicture}
	\end{center}
\caption{%
Integral equation for the point-production amplitude of a particle-dimer system
with total energy $E$ and total momentum $\vec{P}=0$.
Single (double) lines denote particles (dimers), while $g_3$ is the strength of
the point source that creates the particle-dimer pair.
}
\label{fig:3bPointCreation}
\end{figure*}

The point-production amplitude \(\Gamma_\pwl\left(E;p\right)\) is obtained from
the Faddeev equation for the particle-dimer amplitude shown in
Fig.~\ref{fig:3bPointCreation},
\begin{equation}
\Gamma(E;\vec{p})
= \sum_\pwl (2\pwl +1) \Gamma_\pwl\left(E;p\right) P_\pwl(\cos\theta_p) \,,
\end{equation}
by projecting on the partial wave $\pwl$. The same Faddeev equation applies to both the three-boson and three-neutron systems. 
The unified treatment is possible since the running three-body force in the
three-boson system vanishes at a discrete set of ultraviolet cutoffs $\Lambda$
due to its limit-cycle behavior~\cite{Hammer:2000nf}.
We emphasize that this is merely a choice for technical convenience and that
the physics of the three-body force is fully incorporated through the specific
value of the cutoff $\Lambda$.
In the three-neutron case, there is no leading-order three-neutron force due
to Fermi statistics disallowing such an interaction.
Here, the solution is not sensitive to the cutoff $\Lambda$ as long as
$\Lambda$ is sufficiently large.

Apart from the cutoff, both Faddeev equations only differ by a spin-isospin
factor \(\lambda\), which takes the values \(\lambda=1\) for the three-boson and
\(\lambda=-1/2\) for the three-neutron system, respectively.
\begin{widetext}
The equation for the partial-wave projected production amplitude
$\Gamma_\pwl (E;p)$ reads
\begin{equation}\label{eq:Gamma3bInt}
	\begin{aligned}
		\Gamma_\pwl\left(E;p\right) &= g_{3,\pwl} p^{\pwl+2\delta_{l0}\delta_{2\lambda,-1}} 
		+ \lambda\frac{2}{\pi}\int_0^\Lambda \mathrm{d}q ~q^2 Z_{2,\pwl}\left(E;q,p\right) \tau\left(E;q\right) \Gamma_\pwl\left(E;q\right)\,,
  \end{aligned}
  \end{equation}
  where 
  \begin{equation}
      \begin{aligned}
		\tau(E;q) &= \left[ -\frac{1}{a}
  + \sqrt{mE_\text{dimer}(q)} \right]^{-1}\quad\mbox{with} \quad mE_\text{dimer}(q) \equiv \frac{3}{4}{q}^2 - m E - i\varepsilon\,,
   \end{aligned}\label{eq:two-body tmatrix}
  \end{equation}
  and
  \begin{equation}
      \begin{aligned}
  Z_{2,\pwl}\left(E;q,p\right) &=
  \int_{-1}^1 \text{d}x \frac{P_l(x)}{p^2+q^2+pq x-mE-i\varepsilon}\,.
	\end{aligned}\label{eq:one-body propagator}
\end{equation}
\end{widetext}
\(Z_{2,\pwl}\) and \(\tau\) are the two-body interaction kernel and the
dimer propagator, respectively, and the limit $\varepsilon \to 0^+$
is understood at the end of the calculation.
For bosons, note that we work with a specific value of the three-body cutoff $\Lambda$ in
Eq.~\eqref{eq:Gamma3bInt} where the explicit three-body term in the Faddeev
equation vanishes~\cite{Hammer:2000nf}. Since we use a renormalized dimer propagator $\tau$, the two-body cutoff has effectively been taken to infinity.
For fermions, note that the \SW-wave source contains a factor $p^2$ due
to Pauli statistics (see \cref{tab:ConformalSymmetry}).
Equation~\eqref{eq:Gamma3bInt} describes the amplitude for the point
creation of a dimer with two-body scattering length $a$ and a single particle
with angular momentum $\pwl$ (relative to the dimer), where \(g_{3,\pwl}\)
is the strength of the point source.
In the following, this strength will be set to unity for convenience. 
Together with further factors appearing later in the derivation, the source
strength can be understood as a parameter of the theory that determines the
overall normalization and has to be fitted to (experimental) data. The treatment of effective range corrections in the three-neutron case is discussed in Appendix~\ref{sec:effective range corrections}.

From the particle-dimer amplitude \({\Gamma}_\pwl\), the full three-body
amplitude \(\bar{\Gamma}_\pwl\) can be calculated by attaching an external
dimer propagator and a coupling vertex for the break-up into two particles.
Here, a difference between the bosonic and fermionic systems arises because
of the different statistics. 
While \(\bar{\Gamma}_\pwl\) for the three-boson system is given by the
symmetrized sum of all possible permutations of the outgoing particles, the
antisymmetrized sum of the permutations of identical fermions must be used
for the three-neutron system.

\subsection{Three bosons}

First, we consider the three-boson system, using a soft Gaussian
cutoff.\footnote{%
Explicit forms for the effective potentials and more
detailed discussions are given in Ref.~\cite{Dietz:2021haj}.
}
In general, for three distinguishable particles, there are \(3!=6\)
permutations.
However, three of the six permutations are equivalent to the remaining ones,
such that only three terms have to be considered.
\begin{figure*}[htb]
	\begin{center}
		\begin{tikzpicture}[scale=1.0, baseline={([yshift=-.5ex]current bounding box.center)}]
				\node (origin) at (0,0) {};
				\node () at (-1.2, -0.4) {\((E,\vec{0})\)};				
				\node () at (1.3, 1.4) {\(-\frac{\vec{p}}{2} +\vec{k}\)};			
				\node () at (1.5, -0.3) {\(-\frac{\vec{p}}{2}-\vec{k}\)};			
				\node () at (1.5, -1.4) {\(\vec{p}\)};		
				
				\draw  [] (-1.5,0.14) -- (-0.75,0.14);
				\draw  [] (-1.5,0.0) -- (-0.75,0.0);
				\draw  [] (-1.5,-0.14) -- (-0.75,-0.14);

				\draw  [] (0.65,0.4) -- (2.2,1.5);
				\draw  [] (0.75,0.0) -- (2.2,0.0);
				\draw  [] (0.65,-0.4) -- (2.2,-1.5);	
                \node [circle, draw=black, fill=white,inner sep=0mm,minimum size=15mm, line width=0.2mm, pattern=north west lines] () at(origin) {};
                \node [circle, draw=white, fill=white,inner sep=0mm,minimum size=5mm, line width=0.2mm] () at(origin) {};	
				\node () at (0,0) {$\bar{\Gamma}_l$};
		\end{tikzpicture}
		\hspace*{0.5cm}
		=
		\hspace*{0.5cm} 
		\begin{tikzpicture}[scale=1.0, baseline={([yshift=-.5ex]current bounding box.center)}]
				\node (origin) at (0,0) {};
				\node () at (-1.2, -0.4) {\((E,\vec{0})\)};				
				\node () at (1.3, 1.4) {\(-\frac{\vec{p}}{2} +\vec{k}\)};			
				\node () at (1.5, -0.1) {\(-\frac{\vec{p}}{2}-\vec{k}\)};			
				\node () at (1.5, -1.4) {\(\vec{p}\)};		
				
				\draw  [] (-1.5,0.14) -- (-0.75,0.14);
				\draw  [] (-1.5,0.0) -- (-0.75,0.0);
				\draw  [] (-1.5,-0.14) -- (-0.75,-0.14);
			
							
				\draw  [double, double distance=1.4mm] (0.65,0.4) -- (1.2,0.75);	
				\draw  [] (1.2,0.75) -- (2.2,0.0);
				\draw  [] (1.2,0.75) -- (2.2,1.5);
				\node [circle, draw=black, fill=white,inner sep=0mm,minimum size=2mm, line width=0.2mm] () at(1.2,0.75) {};	
				\draw  [] (0.65,-0.4) -- (2.2,-1.5);
				\node [circle, draw=black, fill=white,inner sep=0mm,minimum size=15mm, line width=0.2mm, pattern=north east lines] () at(origin) {};
                \node [circle, draw=white, fill=white,inner sep=0mm,minimum size=5mm, line width=0.2mm] () at(origin) {};	
				\node () at (0,0) {$\Gamma_l$};
		\end{tikzpicture}
		\hspace*{0.5cm}
		+
		\hspace*{0.5cm} 
		\begin{tikzpicture}[scale=1.0, baseline={([yshift=-.5ex]current bounding box.center)}]
				\node (origin) at (0,0) {};

				\node () at (-1.2, -0.4) {\((E,\vec{0})\)};				
				\node () at (1.3, 1.4) {\(-\frac{\vec{p}}{2} -\vec{k}\)};			
				\node () at (1.8, -0.0) {\(\vec{p}\)};			
				\node () at (1.5, -1.6) {\(-\frac{\vec{p}}{2}+\vec{k}\)};

				\draw  [] (-1.5,0.14) -- (-0.75,0.14);
				\draw  [] (-1.5,0.0) -- (-0.75,0.0);
				\draw  [] (-1.5,-0.14) -- (-0.75,-0.14);
			
							
				\draw  [double, double distance=1.4mm] (0.65,0.4) -- (1.2,0.75);	
				\draw  [] (1.2,0.75) -- (2.2,0.0);
				\draw  [] (1.2,0.75) -- (2.2,1.5);
				\node [circle, draw=black, fill=white,inner sep=0mm,minimum size=2mm, line width=0.2mm] () at(1.2,0.75) {};	
				\draw  [] (0.65,-0.4) -- (2.2,-1.5);	
				\node [circle, draw=black, fill=white,inner sep=0mm,minimum size=15mm, line width=0.2mm, pattern=north east lines] () at(origin) {};
                \node [circle, draw=white, fill=white,inner sep=0mm,minimum size=5mm, line width=0.2mm] () at(origin) {};	
				\node () at (0,0) {$\Gamma_l$};
		\end{tikzpicture} \\
		\hspace*{1.5cm}
		+
		\hspace*{0.5cm} 
		\begin{tikzpicture}[scale=1.0, baseline={([yshift=-.5ex]current bounding box.center)}]
				\node (origin) at (0,0) {};

				\node () at (-1.2, -0.4) {\((E,\vec{0})\)};				
				\node () at (1.8, 1.4) {\(\vec{p}\)};			
				\node () at (1.5, -0.1) {\(-\frac{\vec{p}}{2} +\vec{k}\)};			
				\node () at (1.5, -1.6) {\(-\frac{\vec{p}}{2}-\vec{k}\)};

				\draw  [] (-1.5,0.14) -- (-0.75,0.14);
				\draw  [] (-1.5,0.0) -- (-0.75,0.0);
				\draw  [] (-1.5,-0.14) -- (-0.75,-0.14);
			
							
				\draw  [double, double distance=1.4mm] (0.65,0.4) -- (1.2,0.75);	
				\draw  [] (1.2,0.75) -- (2.2,0.0);
				\draw  [] (1.2,0.75) -- (2.2,1.5);
				\node [circle, draw=black, fill=white,inner sep=0mm,minimum size=2mm, line width=0.2mm] () at(1.2,0.75) {};	
				\draw  [] (0.65,-0.4) -- (2.2,-1.5);
				\node [circle, draw=black, fill=white,inner sep=0mm,minimum size=15mm, line width=0.2mm, pattern=north east lines] () at(origin) {};
                \node [circle, draw=white, fill=white,inner sep=0mm,minimum size=5mm, line width=0.2mm] () at(origin) {};	
				\node () at (0,0) {$\Gamma_l$};
		\end{tikzpicture}
	\end{center}
\caption{%
Fully symmetrized point-production amplitude \(\bar{\Gamma}_\pwl\) for three
bosons.
The bosons are produced with total energy $E$ in their center-of-mass, such
that the total momentum is $\vec{P}=0$.
The outgoing bosons are on-shell with the momenta given in the figure.
\label{fig:3bPointCreationGammaBar}
}
\end{figure*}
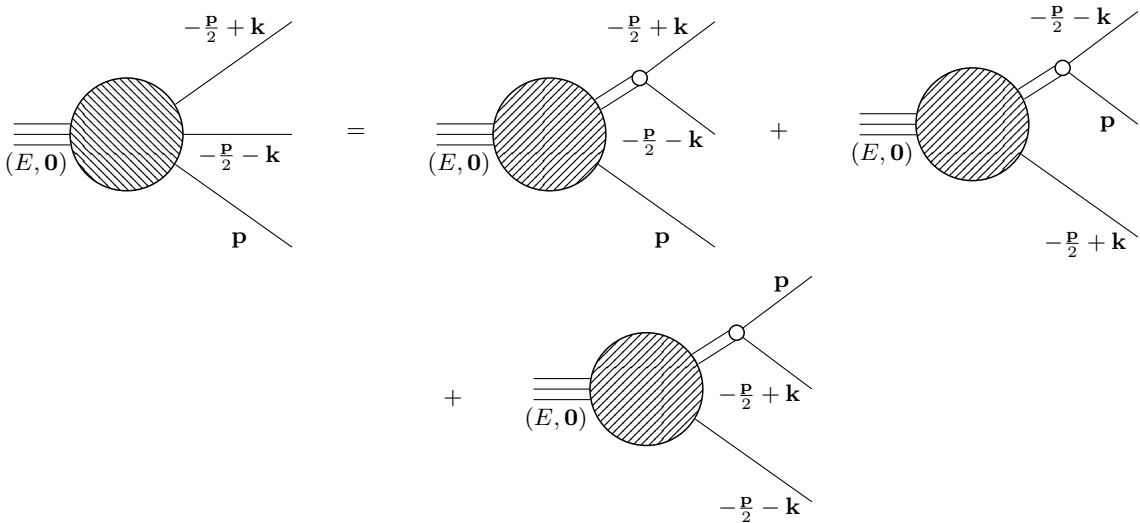
The amplitude is depicted in Fig.~\ref{fig:3bPointCreationGammaBar}, where it is
implicitly assumed that all outgoing particle lines are on-shell.
The corresponding equation for the amplitude \(\bar{\Gamma}_\pwl\) reads
\begin{equation}\label{eq:3bGammaBar}
	\begin{aligned}
		\bar{\Gamma}_\pwl\left(E;\vec{p},
  \vec{k}\right) &= 
		\Gamma_\pwl\left(E;\vec{p}\right) \tau\left(E;-\vec{p}\right) P_\pwl\left( \cos \theta_{\vec{p},\vec{k}} \right) \\ 
		&\phantom{=}+\left(\vec{p}\leftrightarrow-\frac{\vec{p}}{2}+\vec{k}\right)+\left(\vec{p}\leftrightarrow-\frac{\vec{p}}{2}-\vec{k}\right) \\
		\bar{\Gamma}\left(E;\vec{p},
  \vec{k}\right) &\equiv \sum_\pwl \left(2\pwl +1\right) \bar{\Gamma}_\pwl\left(E;\vec{p},
  \vec{k}\right) \,.\,
	\end{aligned}
\end{equation}
where \(\Gamma_\pwl\) is calculated using Eq.~\eqref{eq:Gamma3bInt} with
\(\lambda=1\).
We note that our definition of the partial-wave projected amplitude 
\(\bar{\Gamma}_\pwl\),  directly includes the Legendre polynomial that
describes the angular dependence. 
Finally, the symmetrized amplitude can be used to calculate the
point-production energy distribution \(R(E)\).
It is defined as the phase-space integral of
\(\left| \bar{\Gamma}\left( E; \vec{p},
\vec{k} \right) \right|^2\),
\begin{equation}\label{eq:PointProductionAmplitude}
	\begin{aligned}
		R(E) &= \iint \frac{\mathrm{d}^3k\mathrm{d}^3p}{\left(2\pi\right)^6}\left| \bar{\Gamma} \right|^2
		2\pi \delta
		\left[ E - E_\text{on-shell}\right]\,,
	\end{aligned}
\end{equation}
where the \(\delta\)-function ensures the energy conservation of the three
outgoing on-shell bosons with energy 
$E_\text{on-shell}=E_{\vec{p}} + E_{-\frac{\vec{p}}{2}+\vec{k}} 
+ E_{-\frac{\vec{p}}{2}-\vec{k}}$, with $E_{\vec{p}}\equiv\vec{p}^2/(2m)$.
The energy conserving \(\delta\)-function can be simplified by applying
the variable transformation for \(\delta\) distributions, which results in 
\begin{equation}
	\begin{aligned}
		\delta\Big[ E - E_\text{on-shell} \Big] 
		&= \delta\left[ E -\left( \frac{3p^2}{4m}+\frac{k^2}{m} \right) \right]\\
		&= \frac{m}{2}\frac{\delta\left[ k - \sqrt{mE - \frac{3}{4}p^2} \right]}{\sqrt{mE - \frac{3}{4}p^2}}\,.
	\end{aligned}
\end{equation}
Note that the argument of the square root cannot become negative since
energy conservation limits the value of $p\leq \sqrt{\frac{4}{3}mE}$. 

\subsection{Three neutrons}
\begin{figure*}[htb]
	\begin{center}
			\begin{tikzpicture}[baseline={([yshift=-.5ex]current bounding box.center)}]
				\node (origin) at (0,0) {};

				\node () at (-1.2, -0.4) {\((E,\vec{0})\)};				
				\node () at (1.3, 1.4) {\(-\frac{\vec{p}}{2} +\vec{k}\)};			
				\node () at (1.5, -0.3) {\(-\frac{\vec{p}}{2}-\vec{k}\)};	
				\node () at (1.5, -1.4) {\(\vec{p}\)};		
				
				\node () at (2.2,1.5) {\(\quad \uparrow\)};
				\node () at (2.2,0.0) {\(\quad\downarrow\)};
				\node () at (2.2,-1.5) {\(\quad\uparrow\)};
				
				\draw  [] (-1.5,0.14) -- (-0.75,0.14);
				\draw  [] (-1.5,0.0) -- (-0.75,0.0);
				\draw  [] (-1.5,-0.14) -- (-0.75,-0.14);
			
							
				\draw  [] (0.65,0.4) -- (2.2,1.5);
				\draw  [] (0.75,0.0) -- (2.2,0.0);
				\draw  [] (0.65,-0.4) -- (2.2,-1.5);
				\node [circle, draw=black, fill=white,inner sep=0mm,minimum size=15mm, line width=0.2mm, pattern=north west lines] () at(origin) {};		
                \node [circle, draw=white, fill=white,inner sep=0mm,minimum size=5mm, line width=0.2mm] () at(origin) {};	
				\node () at (0,0) {$\bar{\Gamma}_l$};
		\end{tikzpicture}
		\hspace{0.4cm}
		=
		\hspace{0.4cm}
			\begin{tikzpicture}[baseline={([yshift=-.5ex]current bounding box.center)}]
				\node (origin) at (0,0) {};

				\node () at (-1.2, -0.4) {\((E,\vec{0})\)};				
				\node () at (1.3, 1.4) {\(-\frac{\vec{p}}{2} +\vec{k}\)};			
				\node () at (1.5, -0.3) {\(-\frac{\vec{p}}{2}-\vec{k}\)};			
				\node () at (1.5, -1.4) {\(\vec{p}\)};		
				
				\node () at (2.2,1.5) {\(\quad\uparrow\)};
				\node () at (2.2,0.0) {\(\quad\downarrow\)};
				\node () at (2.2,-1.5) {\(\quad\uparrow\)};
				
				\draw  [] (-1.5,0.14) -- (-0.75,0.14);
				\draw  [] (-1.5,0.0) -- (-0.75,0.0);
				\draw  [] (-1.5,-0.14) -- (-0.75,-0.14);
			
							
				\draw  [double, double distance=1.4mm] (0.65,0.4) -- (1.2,0.75);	
				\draw  [] (1.2,0.75) -- (2.2,0.0);
				\draw  [] (1.2,0.75) -- (2.2,1.5);
				\node [circle, draw=black, fill=white,inner sep=0mm,minimum size=2mm, line width=0.2mm] () at(1.2,0.75) {};	
				\draw  [] (0.65,-0.4) -- (2.2,-1.5);
				\node [circle, draw=black, fill=white,inner sep=0mm,minimum size=15mm, line width=0.2mm, pattern=north east lines] () at(origin) {};
                \node [circle, draw=white, fill=white,inner sep=0mm,minimum size=5mm, line width=0.2mm] () at(origin) {};	
				\node () at (0,0) {$\Gamma_l$};
		\end{tikzpicture}
		\hspace{0.4cm}
		\(-\)
		\hspace{0.4cm}
			\begin{tikzpicture}[baseline={([yshift=-.5ex]current bounding box.center)}]
				\node (origin) at (0,0) {};

				\node () at (-1.2, -0.4) {\((E,\vec{0})\)};				
				\node () at (1.6, 1.4) {\(\vec{p}\)};			
				\node () at (1.5, -0.3) {\(-\frac{\vec{p}}{2}-\vec{k}\)};			
				\node () at (1.2, -1.4) {\(-\frac{\vec{p}}{2}+\vec{k}\)};		
				
				\node () at (2.2,1.5) {\(\quad\uparrow\)};
				\node () at (2.2,0.0) {\(\quad\downarrow\)};
				\node () at (2.2,-1.5) {\(\quad\uparrow\)};
				
				\draw  [] (-1.5,0.14) -- (-0.75,0.14);
				\draw  [] (-1.5,0.0) -- (-0.75,0.0);
				\draw  [] (-1.5,-0.14) -- (-0.75,-0.14);
			
							
				\draw  [double, double distance=1.4mm] (0.65,0.4) -- (1.2,0.75);	
				\draw  [] (1.2,0.75) -- (2.2,0.0);
				\draw  [] (1.2,0.75) -- (2.2,1.5);
				\node [circle, draw=black, fill=white,inner sep=0mm,minimum size=2mm, line width=0.2mm] () at(1.2,0.75) {};	
				\draw  [] (0.65,-0.4) -- (2.2,-1.5);	
				\node [circle, draw=black, fill=white,inner sep=0mm,minimum size=15mm, line width=0.2mm, pattern=north east lines] () at(origin) {};
                \node [circle, draw=white, fill=white,inner sep=0mm,minimum size=5mm, line width=0.2mm] () at(origin) {};	
				\node () at (0,0) {$\Gamma_l$};
		\end{tikzpicture}
	\end{center}
\caption{%
Antisymmetrized point-production amplitude \(\bar{\Gamma}_\pwl\) for three
neutrons.
The neutrons are produced with total energy $E$ in their center-of-mass, such
that the total momentum is $\vec{P}=0$.
The outgoing neutrons are on-shell with the momenta and spins indicated in
the figure.
\label{fig:3nPointCreation}
}
\end{figure*}
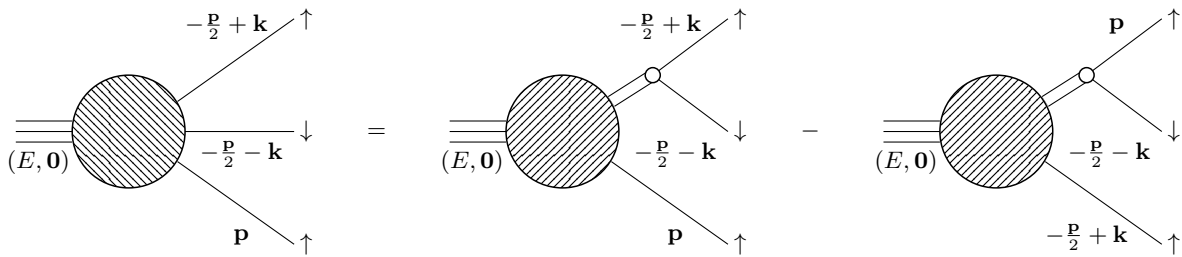

Next, we consider a system of three neutrons.
For this case, we show results obtained with a sharp-cutoff regulator.\footnote{Indeed, we performed the three-neutron calculations using both a soft Gaussian regulator and a sharp cutoff, obtaining consistent results. In the text, we show the sharp-cutoff results.}
According to the Pauli principle, the amplitude now needs to be antisymmetrized
with respect to identical particles.
The two neutrons emerging from the dimer have opposite spins, since identical
fermions do not interact in the \SW\ wave.
Consequently, only two of the three final-state neutrons have the same spin
orientation and the amplitude must be antisymmetric under exchange of these two
neutrons.
The corresponding antisymmetrized amplitude \(\Gamma_\pwl\) is shown
diagrammatically in Fig.~\ref{fig:3nPointCreation}, where again all outgoing
particle lines are on-shell.

The amplitude $\bar{\Gamma}_\pwl$ can be written as 
\begin{equation}
	\begin{aligned}
		\bar{\Gamma}_\pwl\left(E;\vec{p}, 
  \vec{k}\right) &=
		\Gamma_\pwl\left(E;\vec{p}\right) \tau\left(E;-\vec{p}\right)P_\pwl\left(\cos \theta_{\vec{p},\vec{k}}\right)\\
        & \phantom{=}- \left(\vec{p}\leftrightarrow-\frac{\vec{p}}{2}+\vec{k}\right)
		\,,
	\end{aligned}
\end{equation}
where \(\Gamma_\pwl\) is calculated using Eq.~\eqref{eq:Gamma3bInt} with
\(\lambda=-1/2\).

As we did for three bosons, our aim is to calculate the point-production
distribution $R(E)$ from this amplitude.
It is defined analogously to Eq.~\eqref{eq:PointProductionAmplitude}, where
the connection between \(\bar{\Gamma}\) and \(\bar{\Gamma}_\pwl\) is given by 
Eq.~\eqref{eq:3bGammaBar}.
Here, we will consider partial waves up to the \DW\ wave, demonstrating that the \PW-wave contribution dominates.
We will look at the individual contributions of \SW, \PW, and \DW\ waves, as
well as their sum including all interference terms.

The full point-production distribution up to $\pwl=2$ is given by
\begin{equation}\label{eq:R3nFull}
	\begin{aligned}
		R\left(E\right) &= \frac{m}{8\pi^3} \int_0^{\sqrt{4mE/3}} \mathrm{d}p~p^2 \sqrt{ mE - \frac{3}{4}p^2 }  
		\\
        &\times	\int_{-1}^{+1}\mathrm{d}\cos \theta_{\vec{p},\vec{k}} {\left|\sum_{l=0}^2 \left(2\pwl+1\right) \bar{\Gamma}_\pwl\left( E; \vec{p},\vec{k} \right) \right|^2 }\,,
	\end{aligned}
\end{equation}
where $\vert \vec{k}\vert=k=\sqrt{ mE - \frac{3}{4}p^2 }$ due to the $\delta$
distribution in the integral. Furthermore, we assume that matrix element for the initial reaction and the final state interaction of the neutrons factorize and that the energy dependence of the neutron relative energy spectrum at low energies is dominated by the final state interaction (see Ref.~\cite{Hammer:2021zxb} for details).  The point-production distribution can then be related to the differential cross section via $\text{d}\sigma/\text{d}E\sim R(E)$.
To isolate the contribution of individual partial waves, we set the contribution of
the other partial waves to zero. In this case there are no interference terms,
by construction.

\section{Results}
\label{sec:results}

We are now in the position to calculate the point-production amplitudes and
study their behavior. 

\subsection{Three bosons}

We start with the three-boson case, for which we set the boson mass equal to the
neutron mass, to make it easy to compare to the three-neutron results
(discussed below).
Efimov states for negative scattering length $a$ are known to evolve into
resonances as they cross the
three-particle threshold~\cite{Dietz:2021haj,Bringas:2004zz,Deltuva:2020sdd,%
Hyodo:2013zxa,Jonsell:2006xx}.
Thus, it is interesting to investigate the signature of the resonances
in the point-production amplitude.
To calculate the point-production distribution, we solve the integral
equation~\eqref{eq:Gamma3bInt} for \(\Gamma_\pwl\). Here, we use effective potentials with a smooth Gaussian cutoff. For details, see Ref.~\cite{Dietz:2021haj}.
We focus on the case where the third boson is in a relative \SW\ wave,
\(\pwl = 0\), which is the dominant contribution at low energies.
As discussed above, we choose a renormalization scheme where the three-body
force is set to zero and the three-body parameter is varied by changing the
cutoff $\Lambda$ in Eq.~\eqref{eq:Gamma3bInt}.
The integral equation features a three-body branch cut along the positive
real energy axis.
To deal with this, we calculate \(\Gamma_0\) not on the real axis, but
slightly shifted onto the first quadrant of the complex energy plane,
\(E \rightarrow E + \ii \varepsilon\).
In the end, one has to extrapolate to \(\varepsilon=0\), or at least
choose $\varepsilon \ll \text{Re}\ E$.
\begin{figure}[htb]
    \centering
    \includegraphics[width=\linewidth]{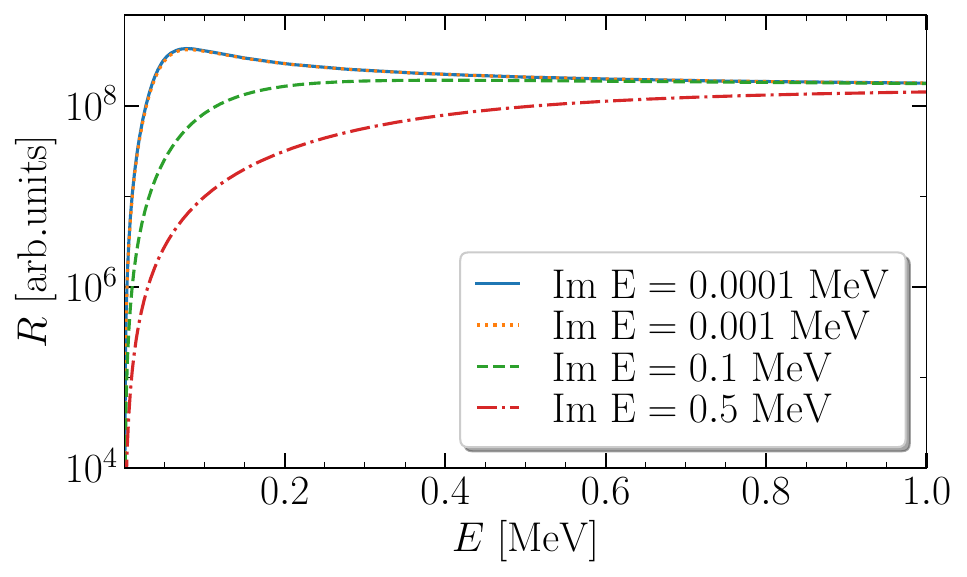}
	\caption{
	Point-production distribution for the three-boson system with
    $\Lambda = 179.9$~fm$^{-1}$ and $a=-12.6$~fm, corresponding to \(a_- / a = 1.5\),
    for different values of the artificial imaginary part of the energy $E$.
    The curves for the two smallest values are indistinguishable.
    In the limit \(\Im E \rightarrow\) 0 the physical amplitude is obtained. 
    \label{fig:Rho3b}
	}
\end{figure}
Figure~\ref{fig:Rho3b} shows the resulting point-production distribution for 
different imaginary parts $\Im E$ of the complex energy. 
In the limit \(\varepsilon = \Im E \to 0\) , the physical result is obtained. 	
The parameter values $\Lambda = 179.9$~fm$^{-1}$ and 
$a=-12.6$~fm were chosen such that \(a_- / a = 1.5\), where \(a_-=-18.9\)~fm is
the scattering length at which the pole trajectory crosses the three-body
branch cut.
The curves for the two smallest values of $\Im E$ are indistinguishable and
show a peak close to threshold.\footnote{Note that we perform all calculations for the ground state trimers. 
We choose the three-body cutoff large enough for finite cutoff corrections to be small. Then all states are to good approximation universal and it does not matter for which states one does the calculation.}
\begin{figure}[htb]
    \centering
    \includegraphics[width=\linewidth]{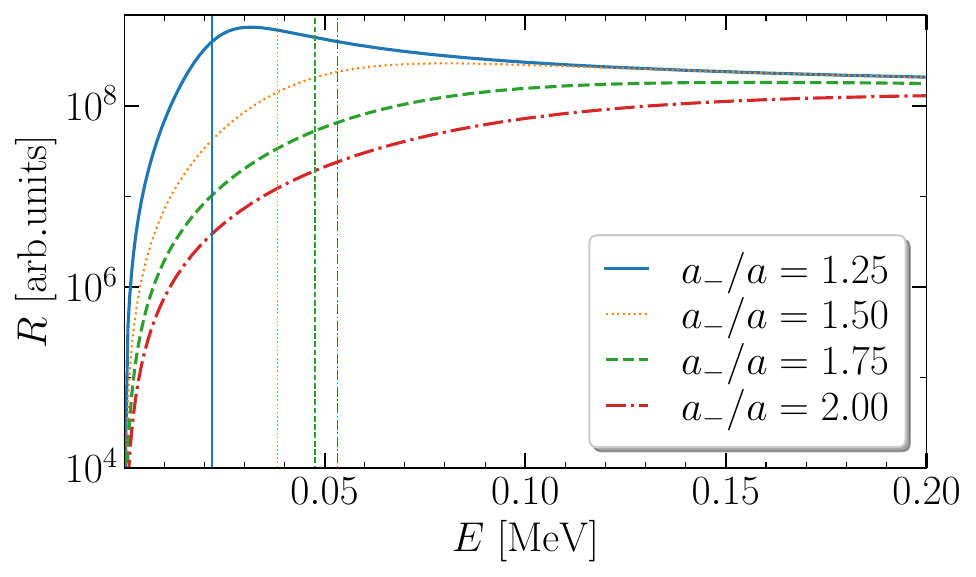}
	\caption{
 Physical point-production distribution in the limit Im E \(\to 0 \)
 for different ratios of \(a_-/a\).
 The parameter values are $\Lambda=179.9$~fm$^{-1}$
and $a=-15.12, -12.6, -10.8,$ and $-9.45$~fm.}
	\label{fig:Rho3bComparison.1}
\end{figure}
To investigate the connection of the peak position to real part of the resonance energy, we show in Fig.~\ref{fig:Rho3bComparison.1} the point-production distribution calculated for different values of \(a_-/a\).
In practice, this was done by keeping $\Lambda=179.9$~fm$^{-1}$
fixed and taking the scattering length as $a=-15.12, -12.6, -10.8,$ and $-9.45$~fm. 
In Fig.~\ref{fig:Rho3bComparison.1}, we show the results for the physical point-production distribution for \(a_-/a = 1.25, 1.5, 1.75,\) and \(2.0\).
The real part of the Efimov resonance energy calculated from an analytical continuation~\cite{Dietz:2021haj} is indicated by a vertical line.

The resonance energies and widths from the analytical continuation for different values of $a_-/a$ are collected in
Tab.~\ref{tab:ResonanceEnergies}.
For small widths $\Gamma$, the peak position of the point-production distribution gives a good estimate
of the real part of the resonance energy.
However, for larger resonance widths the difference between the two positions becomes larger and ultimately the peak disappears altogether.
At $a=-18.9$ fm, \(a_-/a = 1\) and the resonance turns into an Efimov state directly at the three-body threshold, which becomes bound when $|a|$ is increased even further.
\begin{table}[htb]
\begin{center}
\begin{tabular}{c | c | c } 
 \toprule
 \(a_- / a \)  & \(E ~[\mathrm{MeV}]\) & \(\Gamma / 2~[\mathrm{MeV}]\) \\ [0.5ex] 
 \midrule
    \num{1.10} & \num{0.010} & \num{0.003} \\
    \num{1.25} & \num{0.022} & \num{0.014} \\
    \num{1.50} & \num{0.037} & \num{0.040} \\
    \num{1.75} & \num{0.048} & \num{0.074} \\
    \num{2.00} & \num{0.053} & \num{0.113} \\
 \bottomrule
\end{tabular}
 \caption{%
 Resonance energies and widths for different scattering length calculated using the analytical continuation~\cite{Dietz:2021haj}. 
 The transition to a bound state takes place at \(a_- = \qty{{-}18.9}{\femto\meter}\).}
 \label{tab:ResonanceEnergies}
\end{center}
\end{table} 
Our calculations agree with earlier calculations of near-threshold Efimov
resonances in the three-boson system via transition operators on the real energy
axis in Ref.~\cite{Deltuva:2020sdd}.
Moreover, they demonstrate that the point-production distribution contains
information about near-threshold resonances and therefore serve as a benchmark for
our investigation of the three-neutron system. In this context, we note -- consistent with \cref{fig:Rho3bComparison.1,tab:ResonanceEnergies} -- that only sharp resonances with $\Gamma/2\lesssim E$ appear clearly as a peak in the point-production distribution. Broad resonances are inherently difficult to observe and require a dedicated partial wave analysis to identify.

\subsection{Three neutrons}

While the three-boson system at LO is determined by two scales, the scattering
length $a$ and the three-body force (or equivalently the cutoff $\Lambda$), the
three-neutron system features only one scale at leading order, the scattering
length $a$.
In the unitary limit of infinite scattering length, the system exhibits an
extended symmetry, the so-called Schrödinger or nonrelativistic conformal symmetry.
Thus the system can be considered as an approximate realization of an
``unparticle'' or ``unnucleus''~\cite{Hammer:2021zxb}. 

A multineutron system can be described approximately by a field in a nonrelativistic conformal field theory that is characterized by
two parameters, a scaling dimension \(\Delta\) and its
mass~\cite{Nishida:2007pj}. 
The approximate conformal description is applicable to multineutron systems with
center-of-mass energies in the range
\begin{equation}\label{eq:ValidityCFT}
	\begin{aligned}
		\frac{1}{ma^2} \approx \qty{0.1}{\mega\electronvolt}   \ll E \ll \frac{1}{mr^2} \approx \qty{5}{\mega\electronvolt}\,,
	\end{aligned}
\end{equation}
where, in first approximation, the scales $1/(ma^2)$ and $1/(mr^2)$ can be taken to 0 and infinity, respectively.
The region of energies $E\lesssim 1/(ma^2)$ can be calculated in standard
pionless EFT, which includes the effects of finite $a$.
Since we apply pionless EFT at LO in our numerical calculations of
three-neutron systems, the effective range \(r\) effectively takes
the value zero\footnote{%
That is, $r\to0$ in the limit $\Lambda\to\infty$, for the two-body cutoff.
Depending on the particular regularization scheme used to implement the EFT,
at finite $\Lambda$, $r$ may take values that are $\mathcal{O}(1/\Lambda)$,
which can be neglected for sufficiently large $\Lambda$.}.
Therefore, our results show conformal behavior up to arbitrarily high energy,
but this behavior is nonphysical for $E\gtrsim 1/(mr^2)$.

In the conformal region, the energy dependence of the point-production
distribution $R$ is solely determined  by
the scaling dimension of the conformal field (up to an overall normalization). 
Specifically, it scales as~\cite{Hammer:2021zxb} 
\begin{equation}
\label{eq:Rscaling}
	\begin{aligned}
		R(E)  \sim E^{\Delta - \frac{5}{2}} \,,
	\end{aligned}
\end{equation}
where $\Delta$ is the scaling dimension of the 
conformal field representing the multineutron state. 

For very small energies $E\ll 1/(ma^2)$, the system is described by the
scaling dimensions of the non-interacting theory, which are determined by
naive dimensional analysis.
A free field is characterized by the scaling dimension \(\Delta=3/2\).
Thus the energy dependence
of the point-production amplitude for \(N\) particles is given by
Eq.~\eqref{eq:Rscaling}, with $\Delta$ replaced by
\begin{equation}
	\begin{aligned}
		\Delta_\mathrm{free} &= \frac{3N}{2}+\# \vec{\nabla}
        +2\times\#\partial_t
        \,,
	\end{aligned}
\end{equation}
where \(\# \vec{\nabla}\) and \(\# \partial_t\)
are the number of spatial or time derivatives included in the composite
operator \(\mathcal{O}\) that represents the multineutron state. 
An overview of the field operators $\mathcal{O}$ in terms neutron fields
$\psi_\downarrow$ and $\psi_\uparrow$ with the lowest scaling dimensions
$\Delta$ of multineutron systems up to six neutrons is given in
Tab.~\ref{tab:ConformalSymmetry}.

\begin{table}[htb]
\begin{center}
\begin{tabular}{c | c c | c | c | c} 
 \toprule
 \(N\)  & \(S\) & \(L\) & \(\mathcal{O}\)  & \(\Delta\) & \(\Delta_\mathrm{free}\) \\ [0.5ex] 
 \midrule
 \num{2} & \num{0} 	    & \num{0} & \(\psi_\uparrow \psi_\downarrow\) & \num{2} & 3 \\
 \num{3} & 1/2 	& \num{0} & \(\psi_\uparrow \psi_\downarrow \partial_t \psi_\uparrow\) & \num{4.66622} & \num{6.5}   \\
 \num{3} & 1/2 	& \num{1} & \(\psi_\uparrow \psi_\downarrow (\vec{\nabla}_i \psi_\uparrow)\) & \num{4.27272} & \num{5.5}   \\
 \num{3} & 1/2 	& \num{2} & \(\psi_\uparrow \psi_\downarrow \vec{\nabla}_i \vec{\nabla}_j \psi_\uparrow\) & \num{5.60498} & \num{6.5}   \\
 \num{4} & \num{0}		& \num{0} & \(\psi_\uparrow\psi_\downarrow (\vec{\nabla}_i\psi_\uparrow) (\vec{\nabla}_i\psi_\downarrow)\)& \num{5.028} & 8    \\
\num{5} & 1/2		& \num{1} & \(\psi_\uparrow\psi_\downarrow (\vec{\nabla}_i\psi_\uparrow) (\vec{\nabla}_i\psi_\downarrow)(\vec{\nabla}_j\psi_\uparrow) \)
& \num{7.53} & 10.5   \\
\num{6} & \num{0}		& \num{0} & \(\psi_\uparrow\psi_\downarrow ((\vec{\nabla}_i\psi_\uparrow) (\vec{\nabla}_i\psi_\downarrow))^2\)& \num{8.48} & 13   \\
 \bottomrule
\end{tabular}
 \caption{An overview of the operators $\mathcal{O}$ with the lowest scaling
 dimension for different particle numbers $N$ in the vicinity of the strongly
 interacting and free fixed points~\cite{Nishida:2010tm}.
 $S$ and $L$ denote the corresponding spin and orbital angular momentum. Repeated indices are summed over.
 \label{tab:ConformalSymmetry}
 }
\end{center}
\end{table} 

\begin{figure}[htb]
    \centering
    \includegraphics[width=\linewidth]{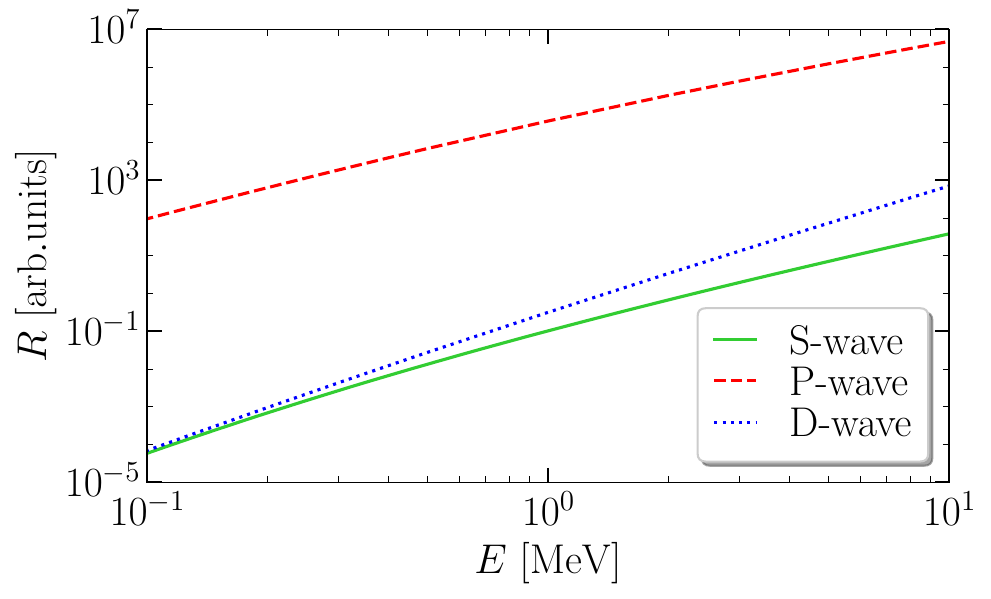}
	\caption{Partial-wave point-production distributions for the three
    partial waves considered.
    For all partial waves an equal strength of the source term,
    \(g_{3,\pwl}=1\) is assumed.
    \label{fig:R3nPWComparison}
    }
\end{figure}

To begin with, one finds that the point-production distributions follow the
predictions of conformal theory for the low and high energy range (details are given 
in Appendix~\ref{sec:Convergence}). However, the partial-wave point-production distributions for
these three partial waves can also be compared with each other
(cf.~Fig.~\ref{fig:R3nPWComparison}). 
Assuming the same strength of the source term \(g_{3,\pwl}=1\) for all partial
waves, the dominant contribution is due to the \PW\ wave.
Under this assumption, the full amplitude up to the \DW\ wave is nearly the same
in comparison to the \PW-wave amplitude, see
Fig.~\ref{fig:R3nPWComparison}.
With this, the central result of this analysis is that, unlike in the
three-boson system, the full three-neutron amplitude shows
\textit{no evidence of any resonance-like structure}. 

We note that our analysis cannot definitely exclude a broad three-neutron resonance that would not lead to a pronounced peak in the point-production distribution. However, our results are in agreement 
with Ref.~\cite{Dietz:2021haj}, which explicitly searched for resonance poles on the unphysical sheet within the same EFT using analytical continuation of the amplitude and did not find any poles. 
Some previous calculations have presented evidence for a three-neutron resonance around 1 MeV
relying on Green's function Monte Carlo calculations with a continuum extrapolation \cite{Gandolfi:2016bth}
and the Gamov no-core shell model \cite{Li:2019pmg}.
The latter paper found a rather large width of about 1 MeV such that the resonance would not show a pronounced structure in the point-production amplitude. The methodology of the first reference was criticized in Ref.~\cite{Deltuva:2019ngx} (see also Ref.~\cite{Gandolfi:2019gmt} for the author's reply).
Our negative conclusion here is also in line with several other recent studies~\cite{Ishikawa:PhysRevC.102.034002,%
Higgins:PhysRevC.103.024004,Deltuva2018:PhysRevC.97.034001}.

However, it is still interesting to study nearly conformal behavior in
the three-neutron system, as discussed above. 
For each partial wave, we introduce a dimensionless function
$\mathcal{R}(x)$ via 
\begin{equation}
    R(E)=c\mathcal{R}(x)E^{\Delta-5/2} \quad \text{for} \quad x=Ema^2 \,,
    \label{eq:dimless r}
\end{equation}
with the conformal scaling dimension $\Delta$ and some constant $c$ such
that $\mathcal{R}(x)\to1$ for $x=Ema^2\to \infty$.
In this way, one can study how rapidly the conformal limit is reached
within our pionless EFT.
\begin{figure}[htb]
    \centering
    \includegraphics[width=\linewidth]{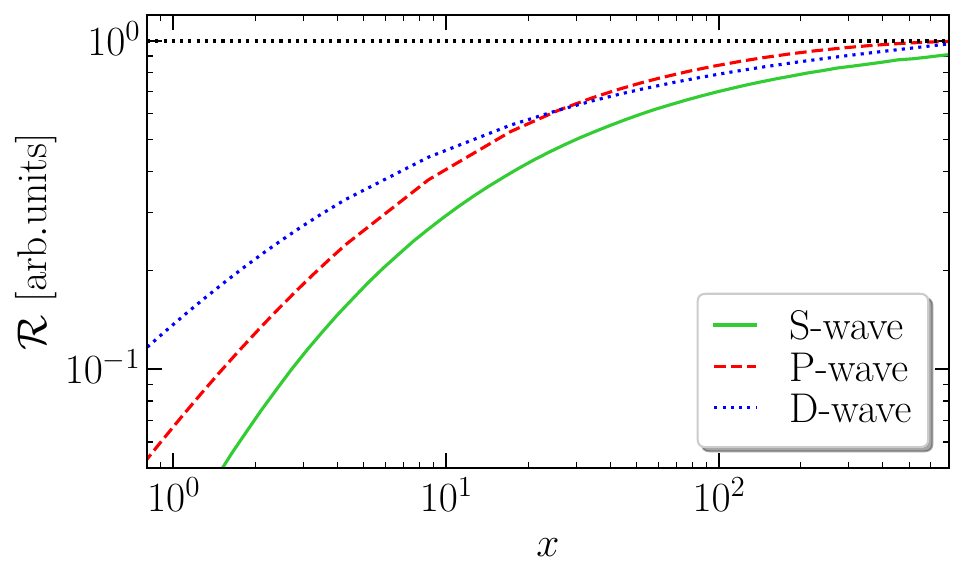}
	\caption{Partial wave point-production distributions for each partial wave considered in its universal form of $\mathcal{R}(x)$ with $x=Ema^2$, defined like in \cref{eq:dimless r}, such that $\mathcal{R}(x)\to1$ for $x\to\infty$. The corresponding coefficients read $c \in[0.386,9.91\cdot10^4,0.723]$ (for \SW, \PW, \DW\ wave).}
	\label{fig:rfunction}
\end{figure}
Investigating the approach of the partial waves towards their respective
conformal limits in \cref{fig:rfunction} shows the following: while the
\PW- and \DW-wave distributions approach the conformal limit at about
$x\simeq 100$, corresponding to about $E\simeq 10$ MeV, the
\SW-wave distribution approaches the conformal limit much slower and
only at higher energies. We further note that compared to \PW\ and
\DW\ wave the \SW\ wave converges more slowly, requiring a careful tuning
of the artificial imaginary part of the energy $\epsilon$ from \cref{eq:two-body tmatrix,eq:one-body propagator} -- for more details see Appendix~\ref{sec:Convergence}.

In Appendix~\ref{sec:effective range corrections}, we calculate the effective range corrections to the point-production distributions treating the effective range non-perturbatively~\cite{Bedaque:1998mb,Bedaque:1997qi}. This amounts to a pionless EFT calculation that is accurate to N$^2$LO, but includes some higher-order range contributions.
The range corrections turn out to be very small. This result is consistent with the perturbative evaluation in conformal field theory, which predicts an exactly vanishing linear range correction to the three-neutron conformal two-point function \cite{chowdhury2024applied}. Interestingly, the linear range correction to the conformal two-point function in the limit of large neutron number also vanishes exactly~\cite{Beane:2025tum}. To our knowledge, the reason for this behavior is not understood.

\section{Comparison with Others}\label{sec:4comparison}

\begin{figure}[htb]
    \centering
    \includegraphics[width=\linewidth]{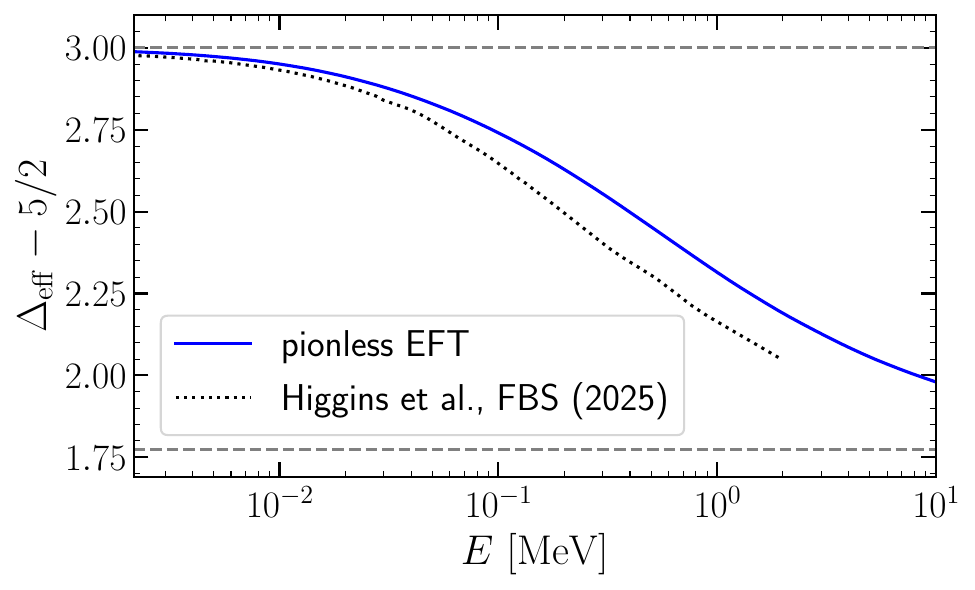}
	\caption{Effective power-law coefficient for three neutron point 
    production.
    Shown are the JWKB results (black dotted) 
    from \textcite{higgins:2025observability} and our pionless EFT results
    (blue solid). For low or high energies the asymptotic values $3$ or
    $1.773$ predicted by conformal field theory (horizontal dashed lines).
    \label{fig:higgens_et_al_comparison}
    }
\end{figure}

We can compare our results with the semiclassical
Jeffries-Wentzel-Kramers-Brillouin (JWKB) approximation results of
\textcite{higgins:2025observability}.
Within their approach  they extracted the effective power-law
coefficient $\Delta_\text{eff}-5/2$ from the relation
$R\propto E^{\Delta_\text{eff}(E)-5/2}$.
In other words, this is just the slope in the log-log plot shown in
our Fig.~\ref{fig:R3nPWComparison}. 
We performed the same kind of analysis for our pionless EFT data, shown in 
Fig.~\ref{fig:higgens_et_al_comparison}.

\begin{figure}[htb]
    \centering
    \includegraphics[width=0.95\linewidth]{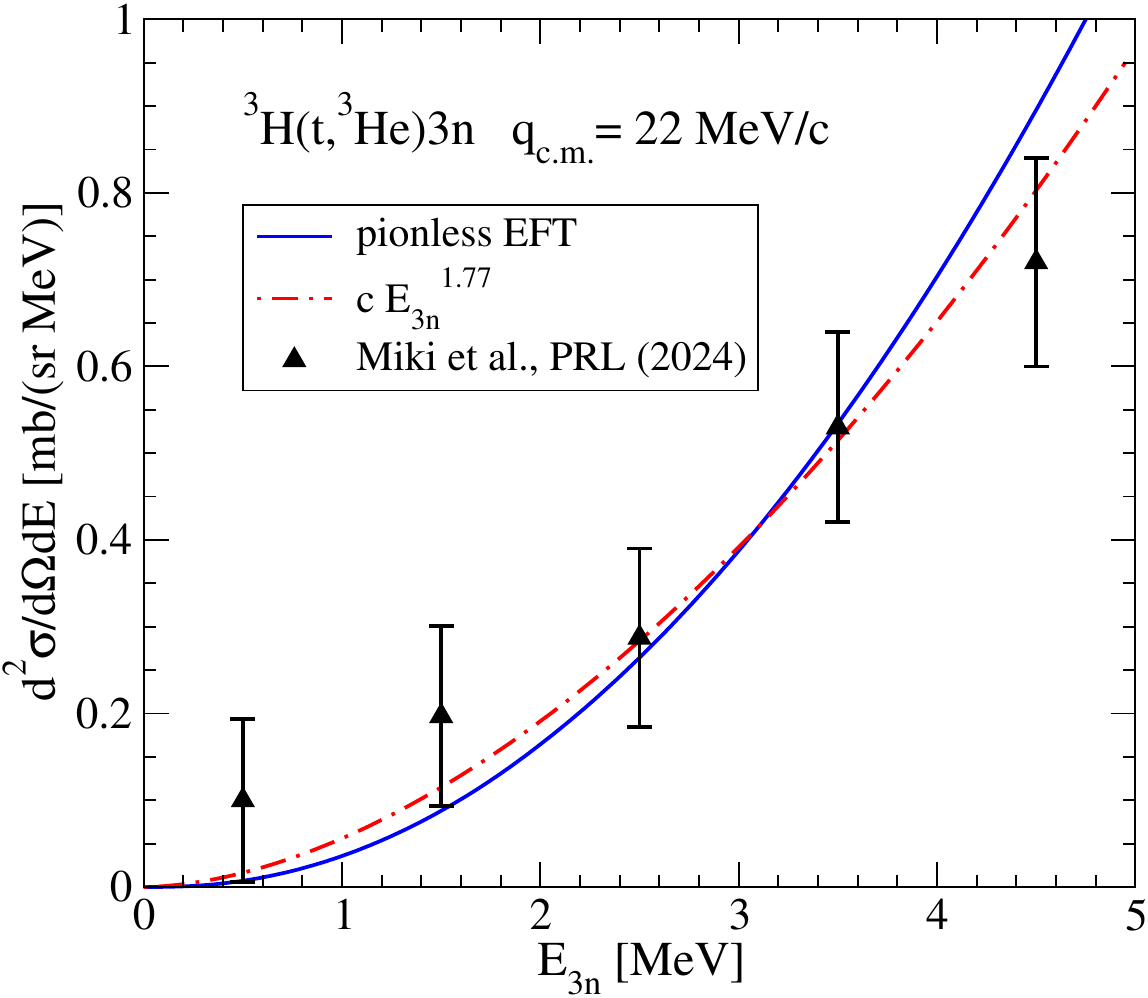}
    \includegraphics[width=0.95\linewidth]{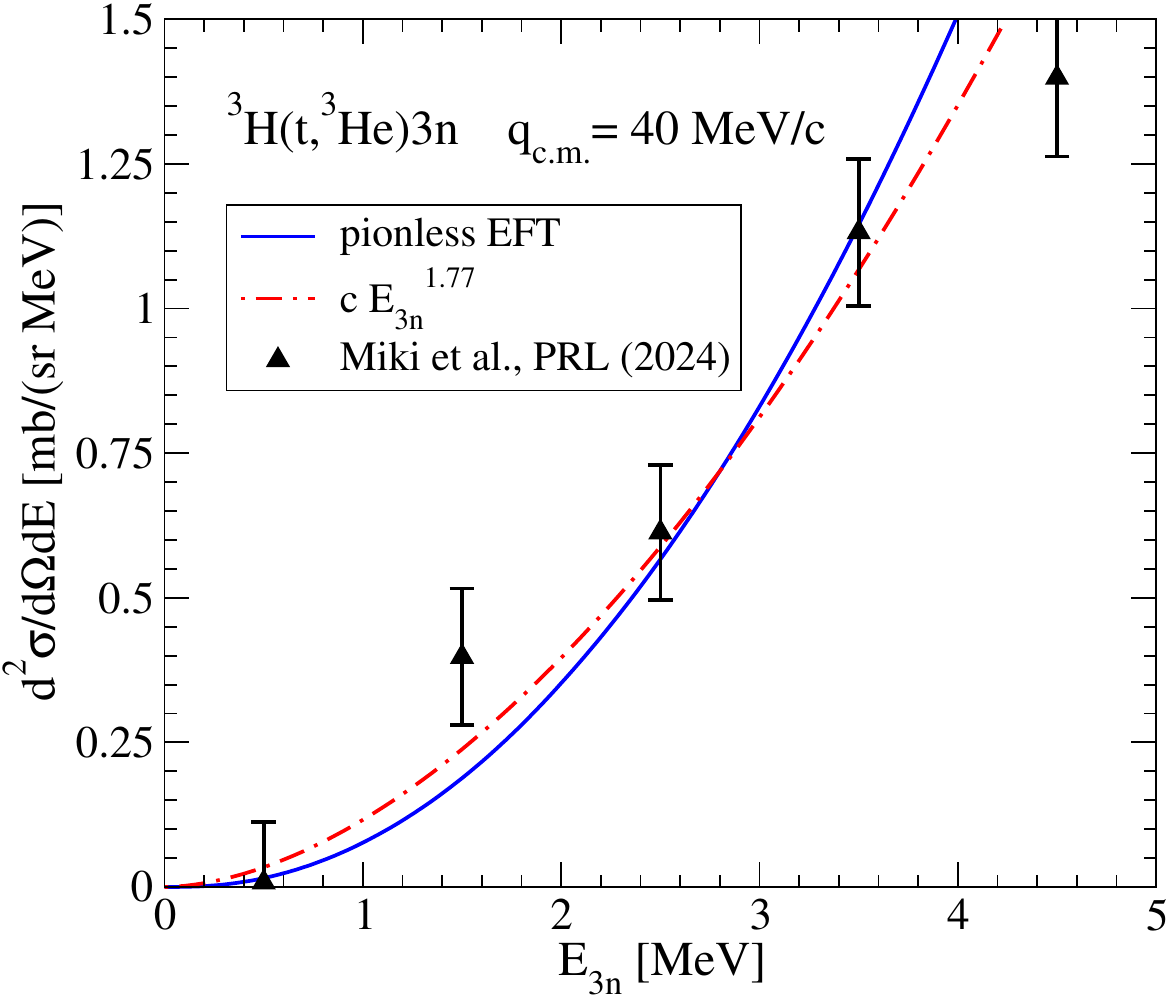}
	\caption{%
    Differential cross section as a function of the relative energy between the
    three neutrons, illustrating the three-neutron continuum.
    Experimental results from \textcite{miki:2024precise} are compared with our
    theoretical results of pionless EFT and conformal field theory.
    The results are shown for average momentum transfer $q_\text{c.m.}=22$ MeV/c
    (upper graph) and $q_\text{c.m.}=40$ MeV/c (lower graph).
    \label{fig:R3n_Miki_et_al}
    }
\end{figure}

In Ref.~\cite{higgins:2025observability}, the authors also plot the effective
exponent for larger values of the scattering length. Compared to our pionless EFT calculation,
their exponent for the physical neutron-neutron scattering length shifts more rapidly as a function of the energy between the asymptotic values
predicted by conformal theory  (see Fig.~\ref{fig:higgens_et_al_comparison}).
Upon closer inspection, one observes that the results of \textcite{higgins:2025observability} for larger values of the scattering lengths overshoot the asymptotic value of $1.773$ at energies around $1$ MeV, suggesting that a similar overshooting may occur for the physical scattering length at larger energies as well (cf.~Fig.~3 of \cite{higgins:2025observability}).
The reason for this is presumably that their hyper-radial potential curve includes some 
higher-order contributions.
This appears to be consistent with our findings including effective range
corrections in Appendix~\ref{sec:effective range corrections}, which indicate that the conformal limit is not reached before the effective range becomes relevant.
In summary, we confirm that the conformal theory asymptotics are not fully
reached within the energy range of interest for the three-neutron continuum.\\ \\
We can furthermore compare our theory calculations with recent experimental
results from \textcite{miki:2024precise}.
These authors studied the low-momentum charge exchange
$^3\text{H}(\text{t},^3\text{He})3n$ reaction to investigate the
three-neutron continuum at intermediate energies.
In Fig.~\ref{fig:R3n_Miki_et_al} we compare our predictions from pionless
EFT and conformal field theory with the actual experimental data from
Ref.~\cite{miki:2024precise} for momentum transfers $q_\text{c.m.}=22,40$ MeV to the three-neutron system. For these low momentum transfers, the three-neutron system is left almost undisturbed in the reaction.

Overall, we see both qualitative and quantitative agreement between pionless EFT,
conformal theory, and the experimental results. The experimental data are not precise enough to differentiate with the conformal prediction and the pionless EFT calculation which includes the effect of the finite scattering length.
Moreover, there is no evidence for a resonance-like structure in the three-neutron continuum in the data of~\cite{miki:2024precise}.
This is in agreement with the pionless EFT and conformal theory results obtained here. The conformal theory also excludes a resonance-like structure in the low-energy four-neutron continuum in agreement with most other theoretical calculations \cite{Marques:2021mqf,Lazauskas:2022mvq}.
Most recently, \textcite{Meißner26:wu2026searching} investigated the
tetraneutron system on the lattice and also reported no indication of a resonance.
For us, this motivates further studies focused on the origin of the structure
observed in the tetraneutron experiment of Ref.~\cite{Duer:2022ehf}.

\section{Summary and outlook}
\label{sec:conc}

We studied the short-range production of three particles characterized by a
large scattering length using pionless EFT.
We introduced a point-production amplitude, which we then solved employing
the Faddeev formalism.
Due to the formal similarity between both systems, we solved the Faddeev
equations for both three identical bosons and three identical fermions, with
particular focus on the potential appearance of resonance-like structures in
the three-neutron system.
In doing so, we compared our pionless EFT results with the findings of
conformal field theory, which are valid in  the unitary limit $1/a\to 0$ and $r\to 0$,
corresponding to the high energy regime of pionless EFT at leading order.
By analyzing each individual partial wave, we confirmed the presence of a
resonance in the three-boson system, as expected from previous
analyses~\cite{Bringas:2004zz,Jonsell:2006xx,Hyodo:2013zxa,Deltuva:2020sdd,%
Dietz:2021haj}.
In contrast, no resonance-like structure was observed in the three-neutron
system, again in line with previous studies~\cite{Ishikawa:PhysRevC.102.034002,%
Higgins:PhysRevC.103.024004,Deltuva2018:PhysRevC.97.034001,Dietz:2021haj}.
Upon examining \SW, \PW\ and \DW\ waves for three neutrons, we found that for
natural values of the production strengths, $g_{3,l}\simeq1$, the
\PW-wave contribution is dominant.
We also carried out a non-perturbative calculation of effective range corrections to the point-production distributions in Appendix~\ref{sec:effective range corrections} amounting to a pionless EFT calculation accurate to N$^2$LO.
The range corrections are very small, consistent with the their perturbative evaluation in conformal field theory~\cite{chowdhury2024applied} and the result for large neutron numbers~\cite{Beane:2025tum}.

In light of recent and forthcoming tetraneutron experiments suggesting a
structure in the final tetraneutron state~\cite{Duer:2022ehf}, an analysis
of the four-neutron system, similar to the one performed here for three neutrons,
would be very interesting.
In future work, we plan to do this using Faddeev-Yakubovsky formalism.
From the perspective of conformal field theory, as discussed within this work,
no such resonance structure is expected.
This is in line with the findings of Ref.~\cite{Lazauskas:2022mvq}, who argue
that the resonance-like structure
reported in Ref.~\cite{Duer:2022ehf} is caused by the initial four-neutron
configuration originating from the structure of \isotope[8]{He}.
\textcite{Zhang:2025uin} recently found similar four-neutron configurations in nuclear lattice EFT.

\begin{acknowledgments}
This work was supported in part by the Deutsche Forschungsgemeinschaft (DFG,
German Research Foundation) -- Project ID 279384907 -- SFB 1245 
(T.G.B., S.D. and H.W.H.), by the BMFTR Contract No.~05P24RDB (H.W.H.), by the U.S.\ National Science Foundation
under Grant No.~PHY--2044632 (S.K.), and by the U.S.\ Department of Energy
under Grant No.\ DE-FG02-13ER41958 (D.T.S.).
This material is based upon work supported by the U.S.\ Department of Energy,
Office of Science, Office of Nuclear Physics, under the FRIB Theory Alliance
Award DE-SC0013617 (S.K.).
Computational resources for parts of this work were provided by the
high-performance computing cluster operated by North Carolina State University,
as well as by the Jülich Supercomputing Centre,
Forschungszentrum Jülich, Germany.
\end{acknowledgments}

\section*{Data Availability}
The data that support the findings of this article are not publicly available. The data are available from the authors upon reasonable request.

\appendix

\section{Effective range corrections}
\label{sec:effective range corrections}

In this section, we incorporate the effective range $r=2.83$~fm to obtain
and analyze the effective range corrections in the three-neutron case for both \SW\ and \PW\ wave.

To introduce the effective range in our theory, we adept the two-body
T-matrix from Eq.~\ref{eq:two-body tmatrix} as follows
\begin{equation}
      \begin{aligned}
		\tau_\text{N$^2$LO}(E;q)^{-1} &= \tau(E;q)^{-1}-\frac{r}{2} mE_\text{dimer}(q)\,,
  \\ \mbox{with}& \quad mE_\text{dimer}(q) \equiv \frac{3}{4}{q}^2 - m E - i\varepsilon\,.\label{eq:N2LO two-body tmatrix}
   \end{aligned}
  \end{equation}
By proceeding this way, which corresponds to a resummation of the terms associated
with $C_2$ in Eq.~\eqref{eq:2BInt}, we include all contributions up to N$^2$LO
contributions and even some of higher order~\cite{Bedaque:1998mb,Bedaque:1997qi}.
We note that in principle such corrections should be treated in perturbation theory, as dictated by the power counting.
Moreover,
in a calculation strictly based on a potential as in Eq.~\eqref{eq:2BInt}, one
runs into the so-called ``Wigner bound'' that would limit the maximum value of our
cutoff $\Lambda$~\cite{Wigner:1955zz,Phillips:1996ae,Hammer:2009zh}.
However, in the diagrammatic formulation with dimer/dineutron fields, it is possible
to avoid this bound (by implementing the range correction effectively as an energy-dependent potential).
Since the resulting integral equation for three neutrons is well behaved~\cite{Bedaque:1998mb,Bedaque:1997qi}, we find it useful to use the resummed
approach to estimate the influence of effective-range corrections.
Their strictly perturbative treatment will be considered in future work.
Aside from modifying the two-body
T-matrix according to Eq.~\eqref{eq:N2LO two-body tmatrix}, the procedure to determine the production distribution remains the same. We note in passing that in the three-boson case such a treatment in the resummed approach is not possible due to spurious poles in the dimer propagator.

\begin{figure}[htb]
    \centering
    \includegraphics[width=\linewidth]{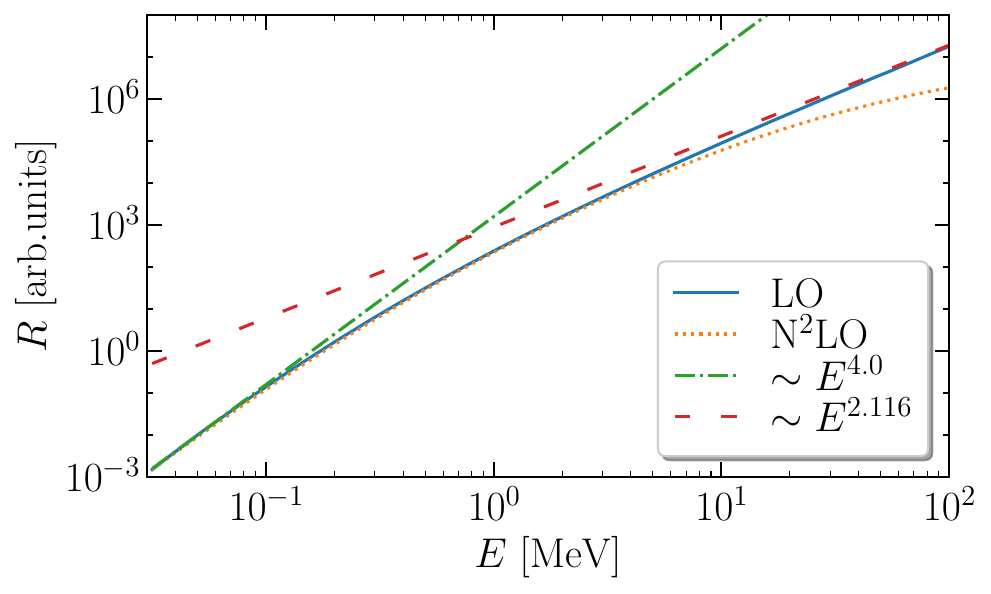}
    \includegraphics[width=\linewidth]{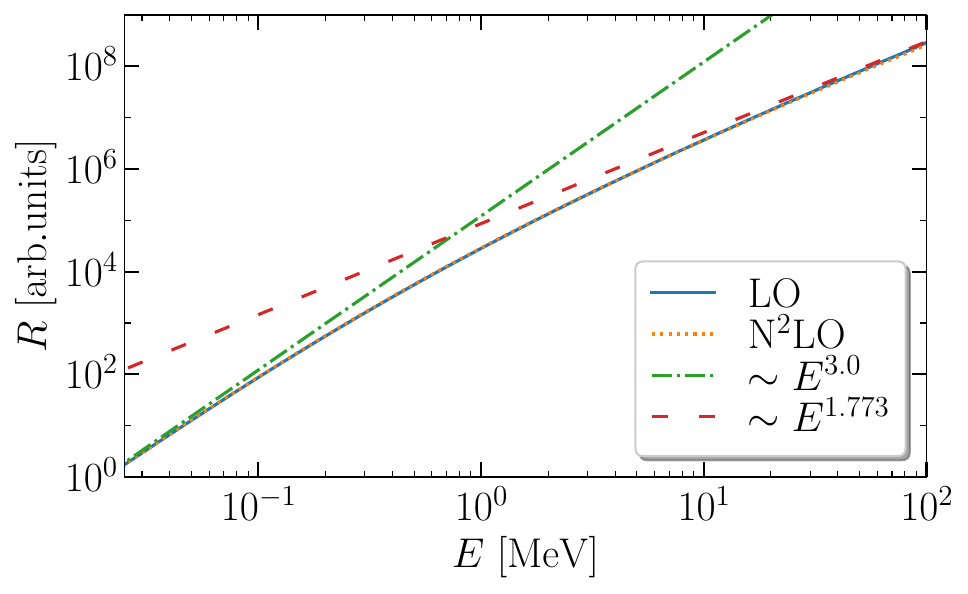}
	\caption{Partial-wave point-production distribution $R$ for the three-neutron
    system for pionless EFT in LO (blue solid line) and N$^2$LO (orange dotted line),
    with the third neutron in a relative \SW\ and \PW\ wave, are shown in the upper
    and lower panel, respectively.
    The small or high energy results can be compared to the predictions for free
    particles (green dash-dotted line) or nonrelativistic conformal field theory
    (red loosely dashed line). We see very good agreement for both low and high
    energies at LO and N$^2$LO, with expected deviations at higher energies.
    \label{fig:R3nWavesN2LO}
    }
\end{figure}

In \cref{fig:R3nWavesN2LO} we plot \SW- and \PW-wave results in LO and N$^2$LO.
The LO and N$^2$LO results are consistent at low energies -- as expected.
However, as energy increases, relative deviations emerge -- at
$1/(mr^2)\approx5$~MeV about 20\% for the \SW\ and 0.4\% for the \PW\ wave,
respectively.
The \SW\ wave is consistent with our rough estimate of $|r/a|\simeq15\%$, but the effective range contribution to the \PW\ wave is strongly suppressed.
Notably, the effective range $r=2.83$ fm plays a more significant role in
the \SW-wave than in the \PW-wave results.
This means up to N$^2$LO the \PW\ wave still reaches the conformal limit,
while the \SW\ wave does not -- or only approximately so at about 10~MeV.
These results are in line with the findings of \textcite{chowdhury2024applied}, which demonstrated that the first-order
effective range correction to conformal theory in the three neutron case vanishes exactly, making the effect of the effective range rather small. 
Moreover, in the limit of large neutron number the linear range correction also vanishes exactly~\cite{Beane:2025tum}.
In fact, our numerical results are consistent with the subleading $r^2$ correction vanishing as well. In contrast, the $r^2$ correction in the expansion for large neutron number turns out to be finite \cite{Beane:2025tum}. This provides further motivation for a strictly perturbative calculation in our approach.

Furthermore, the fact that the \SW\ wave does not reach the conformal limit at N$^2$LO is consistent
with the findings of \textcite{higgins:2025observability},
but in their calculation the same was true  
for the \PW\ wave as well, presumably since higher
orders of the EFT expansion were effectively included.

\section{Consistency with conformal predictions}\label{sec:Convergence}

In this section, we briefly discuss the less convergent nature of \SW-wave
point-production, followed by a consistency check between conformal theory
predictions and the low- and high-energy regime of our leading-order
pionless EFT results.

Let us first consider the far off-shell limit $p^2\gg E$ of Eq.\eqref{eq:Gamma3bInt}
and apply a power-law ansatz for the point-production amplitude.
This yields the following algebraic equation for the conformal scaling
dimension of the \SW-wave channel~\cite{Braaten:2004rn}
\begin{align}
    \Gamma_0(E,p)&\sim p^{\Delta-7/2}=p^2-\dfrac{2}{\pi\sqrt{3}}p^{\Delta-7/2}\nonumber\\&\times\int_0^\infty \text{d}x x^{\Delta-7/2}\ln\left[\dfrac{1+x+x^2}{1-x+x^2}\right]\ .
\end{align}
Taking a closer look, this integrals scales as
$\Lambda^{\Delta-7/2}$ using a sharp cutoff.
Further, focusing on the asymptotic equation, which can be solved via
identification as an (extended) Mellin transform.
This leads to
\begin{align}
    1=-\dfrac{4}{\sqrt{3}}\dfrac{\sin(s\pi/6)}{s\cos(s\pi/2)}, \text{ with } \Delta=s+5/2\ .
\end{align}
One solution, we already expect from our conformal field theory findings,
reads $\Delta=4.66622$.
However, for $\Delta=4.66622$ the integral above does diverge like
$\Lambda^{1.16622}$.
In contrast, applying the same steps to higher partial waves yields an
integral that is at least less divergent or even convergent.
This owes to the presence of the angular-momentum potential barrier.
One can easily verify this for the \PW\ and \DW\ wave. 
In these cases, the steps outlined above lead to an integral that scales
like $\Lambda^{\Delta - 9/2}$ or $\Lambda^{\Delta - 11/2}$, respectively.
For the conformal solutions $\Delta = 4.27272$ and $\Delta = 5.60498$, this
corresponds to $\Lambda^{-0.22728}$ and $\Lambda^{0.10498}$, respectively.
Based on this insight, we suggest that \PW- and \DW-wave calculations are
``more convergent'' than the \SW-wave one.
However, it should be noted that we observed this behavior in the far
off-shell regime.
Taken together, we conclude that the diverging asymptotic solution of the
\SW-wave integral equation likewise signals the numerically observed
instability, which occurs for \SW\ wave but not the other partial waves.
\begin{figure}[htb]
    \centering
    \includegraphics[width=\linewidth]{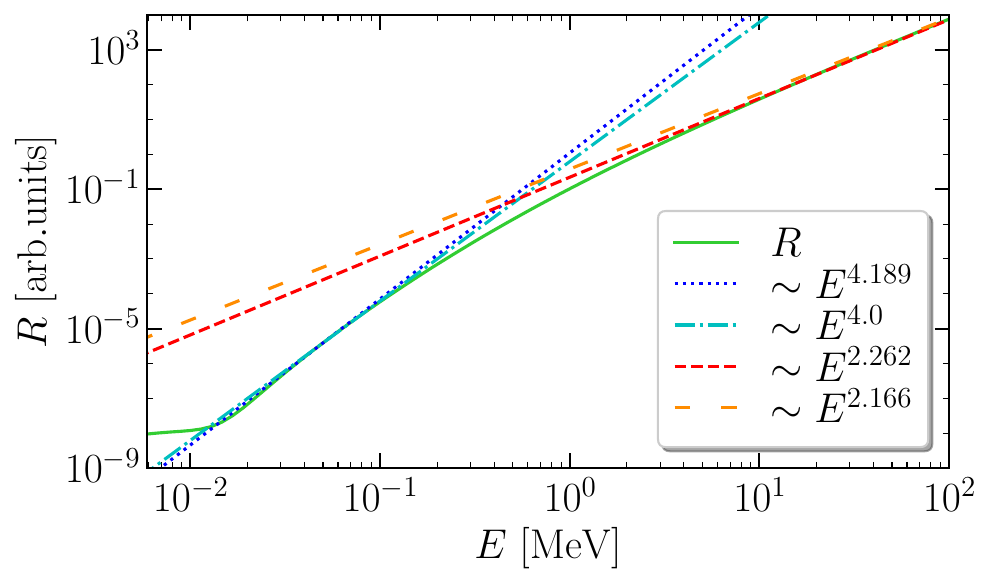}
    \includegraphics[width=\linewidth]{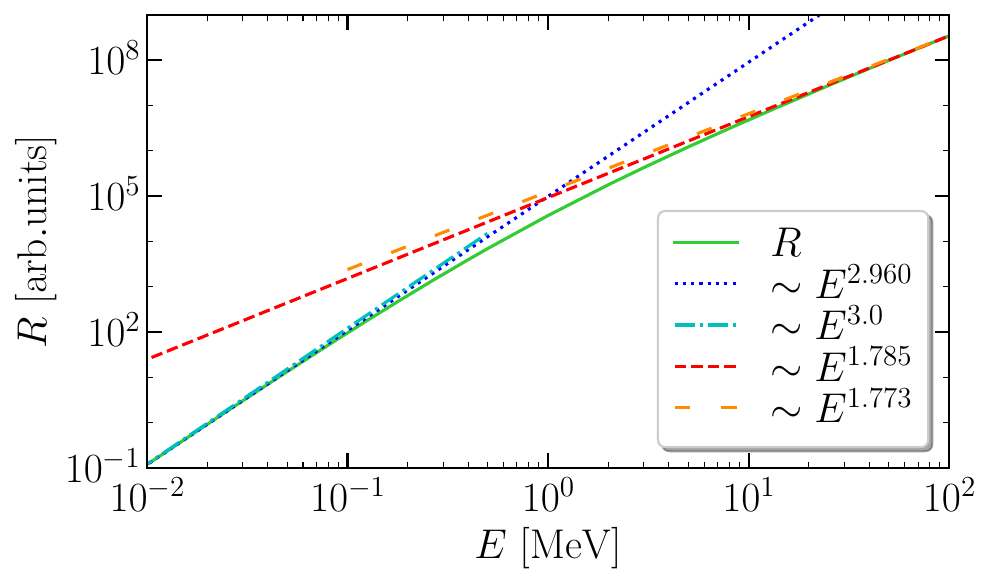}
    \includegraphics[width=\linewidth]{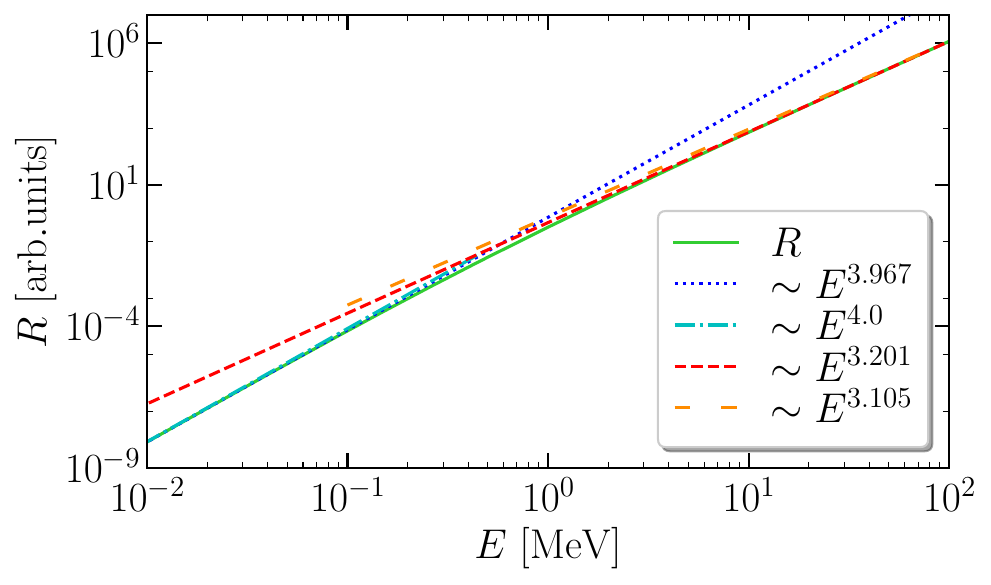}
	\caption{Partial-wave point-production distribution $R$ for the three-neutron
    system, with the third neutron in a relative \SW, \PW, and \DW\ wave,
    are shown in the upper, middle and lower panel, respectively.
    The results are fitted by a function \(R(E)=aE^b\) at small
    (E=\SI{0}{\mega\electronvolt}, blue dotted line) as well as large
    energies (E=\SI{100}{\mega\electronvolt}, red dashed line).
    The small energy fit can be compared to the predictions for free
    particles (cyan dash-dotted line) as well as the fit at large
    energies to the predictions by nonrelativistic conformal field
    theory (orange loosely dashed line).
    Both show a very good agreement between conformal theory and calculation.
    \label{fig:R3nWaves}
    }
\end{figure}
For our calculations, with $\Im{E}=10^{-4}$ MeV, shown in \cref{fig:R3nWaves}, this instability is evident for energies lower
than $E\simeq 10^{-2}$ MeV.
To extract the correct low-energy behavior, the imaginary part of the
energy $\epsilon$ must be tuned much more carefully and requires a much
finer momentum mesh to achieve convergence than in the case of higher
partial waves.\\ \\
However, for fermion-dimer scattering the inhomogeneous term is not
$\simeq p^2$ as in point-production, but instead suppressed like
$\simeq p^{-2}$ at large momenta.
As a consequence, each term of the perturbative Borne series converges,
which is not the case for point-production.
This again highlights the role of the special driving term in the
point-production Faddeev equation.
Moreover, comparing the bosonic case, with a driving term scaling as
$\simeq p^0$, to the fermionic case, where it scales as $\simeq p^2$,
we find -- as expected -- that the bosonic case exhibits better
convergence.\\ \\
In summary, we found differences in the convergence behavior of
different partial waves in our point-production calculations, which
can be traced back to the general structure of the three-body Faddeev
equation, with particular emphasis on the role of the driving term.\\
Referring again to \cref{fig:R3nWaves}, we now compare our EFT results with
the conformal predictions for the first partial waves, namely
\SW, \PW\text{ }and \DW\ waves.
For all three partial-wave channels the results for the point-production
distribution agree reasonably well with these predictions.
Of course this is as one would expect since the low energy regime
$E\ll1/(ma^2)$ corresponds to the free limit $a\to0$ while the high energy
regime $E\gg 1/(ma^2)$ corresponds to the unitary/conformal limit $|a|\to\infty$,
noting that $r=0$~fm within pionless EFT at leading order.

\bibliography{refs}

\end{document}